\documentclass[aps,prx,reprint,superscriptaddress]{revtex4-2}

\usepackage{graphicx}	

\usepackage{hyperref}

\usepackage{amsmath}
\newcommand{\angstrom}{\textup{\AA}}

\usepackage{soul}
\usepackage{color}
\usepackage{xcolor}

\newcommand{\Ag}[1]{$\textrm{A}_{1\textrm{g}}^{#1}$}
\newcommand{\Au}[1]{$\textrm{A}_{1\textrm{u}}^{#1}$}
\newcommand{\Eg}[1]{$\textrm{E}_{\textrm{g}}^{#1}$}
\newcommand{\Eu}[1]{$\textrm{E}_{\textrm{u}}^{#1}$}
\newcommand{\BT}{Bi$_2$Te$_3$}
\newcommand{\BS}{Bi$_2$Se$_3$}
\newcommand{\sglcol}{1\columnwidth}
\newcommand{\dblcol}{2\columnwidth}
\newcommand{\EF}{$E_{\textrm{F}}$}
\newcommand{\one}{{\ensuremath{\pmb{k}}}}

\begin{document}

\title{Influence of local symmetry on lattice dynamics coupled to topological surface states}

\author{Jonathan A. Sobota}
\affiliation{Stanford Institute for Materials and Energy Sciences, SLAC National Accelerator Laboratory, Menlo Park, California 94025, USA}

\author{Samuel W. Teitelbaum}
\affiliation{Stanford PULSE Institute, SLAC National Accelerator Laboratory, Menlo Park, California 94025, USA}
\affiliation{Department of Physics, Arizona State University, Tempe, Arizona 85281, USA}

\author{Yijing Huang}
\affiliation{Stanford PULSE Institute, SLAC National Accelerator Laboratory, Menlo Park, California 94025, USA}
\affiliation{Department of Applied Physics, Stanford University, Stanford, California 94305, USA}

\author{José D. Querales-Flores}
\affiliation{Tyndall National Institute, Lee Maltings, Dyke Parade, Cork T12 R5CP, Ireland}

\author{Robert Power}
\affiliation{Department of Physics, University College Cork, College Road, Cork T12 K8AF, Ireland}

\author{Meabh Allen}
\affiliation{Department of Physics, University College Cork, College Road, Cork T12 K8AF, Ireland}

\author{Costel R. Rotundu}
\affiliation{Stanford Institute for Materials and Energy Sciences, SLAC National Accelerator Laboratory, Menlo Park, California 94025, USA}

\author{Trevor P. Bailey}
\affiliation{Department of Physics, University of Michigan, Ann Arbor, Michigan 48109, USA}

\author{Ctirad Uher}
\affiliation{Department of Physics, University of Michigan, Ann Arbor, Michigan 48109, USA}

\author{Tom Henighan}
\affiliation{Stanford PULSE Institute, SLAC National Accelerator Laboratory, Menlo Park, California 94025, USA}
\affiliation{Department of Applied Physics, Stanford University, Stanford, California 94305, USA}

\author{Mason Jiang}
\affiliation{Stanford PULSE Institute, SLAC National Accelerator Laboratory, Menlo Park, California 94025, USA}
\affiliation{Department of Applied Physics, Stanford University, Stanford, California 94305, USA}

\author{Diling Zhu}
\affiliation{Linac Coherent Light Source, SLAC National Accelerator Laboratory, Menlo Park, California 94025, USA}

\author{Matthieu Chollet}
\affiliation{Linac Coherent Light Source, SLAC National Accelerator Laboratory, Menlo Park, California 94025, USA}

\author{Takahiro Sato}
\affiliation{Linac Coherent Light Source, SLAC National Accelerator Laboratory, Menlo Park, California 94025, USA}

\author{Mariano Trigo}
\affiliation{Stanford PULSE Institute, SLAC National Accelerator Laboratory, Menlo Park, California 94025, USA}
\affiliation{Stanford Institute for Materials and Energy Sciences, SLAC National Accelerator Laboratory, Menlo Park, California 94025, USA}

\author{\'Eamonn D. Murray}
\affiliation{Tyndall National Institute, Lee Maltings, Dyke Parade, Cork T12 R5CP, Ireland}

\author{Ivana Savić}
\affiliation{Tyndall National Institute, Lee Maltings, Dyke Parade, Cork T12 R5CP, Ireland}

\author{Patrick S. Kirchmann}
\affiliation{Stanford Institute for Materials and Energy Sciences, SLAC National Accelerator Laboratory, Menlo Park, California 94025, USA}

\author{Stephen Fahy}
\affiliation{Department of Physics, University College Cork, College Road, Cork T12 K8AF, Ireland}
\affiliation{Tyndall National Institute, Lee Maltings, Dyke Parade, Cork T12 R5CP, Ireland}

\author{David. A. Reis}
\affiliation{Stanford PULSE Institute, SLAC National Accelerator Laboratory, Menlo Park, California 94025, USA}
\affiliation{Stanford Institute for Materials and Energy Sciences, SLAC National Accelerator Laboratory, Menlo Park, California 94025, USA}
\affiliation{Department of Applied Physics, Stanford University, Stanford, California 94305, USA}
\affiliation{Department of Photon Science, Stanford University, Stanford, California 94305, USA}

\author{Zhi-Xun Shen}
\affiliation{Stanford Institute for Materials and Energy Sciences, SLAC National Accelerator Laboratory, Menlo Park, California 94025, USA}
\affiliation{Department of Applied Physics, Stanford University, Stanford, California 94305, USA}

\date{\today}

\begin{abstract}
We investigate coupled electron-lattice dynamics in the topological insulator \BT{} with time-resolved photoemission and time-resolved x-ray diffraction. It is well established that coherent phonons can be launched by optical excitation, but selection rules  generally restrict these modes to zone-center wavevectors and Raman-active branches.    We find that the topological surface state couples to   additional modes, including a continuum of surface-projected bulk modes from both Raman- and infrared-branches, with possible contributions from surface-localized modes when they exist.   Our calculations show that   this surface vibrational spectrum occurs    naturally as a consequence of the translational and inversion symmetries broken at the surface, without requiring the   splitting-off of  surface-localized   phonon modes. The generality of this result suggests that   coherent phonon spectra are    useful by providing unique fingerprints for identifying surface states in more controversial materials.  These effects may also expand the phase space for tailoring surface state wavefunctions via ultrafast optical excitation.
\end{abstract}

\maketitle

\section{Introduction}

A goal of condensed matter physics is to tailor electronic states on demand using ultrafast pulses of light \cite{Basov_2017_towards}. Topological materials provide an appealing platform for this approach, with their non-trivial surface states offering potential applications ranging from spintronics to quantum computing \cite{Fu_2007_topological,Zhang_2009_topological,Hasan_2010_colloquium,Qi_2011_topological}. Strategies range from ``Floquet engineering,'' which employs  the periodic electric field within the pulse \cite{Yao_2007_optical,Lindner_2011_Floquet,Kitagawa_2011_transport,Bukov_2015_universal,Weber_2021_ultrafast}, to ``lattice engineering,'' in which electronic wavefunctions are modified through light-induced structural distortions    \cite{Kim_2015_topological,Moller_2017_type,Wang_2017_phonon,Weber_2018_using,Sie_2109_an,Vaswani_2020_light,Chaudhary_2020_phonon,Weber_2021_ultrafast,Luo_2021_a}. For the latter, it is broadly relevant to understand the pathways through which  ultrafast pulses interact with the crystal lattice and the resulting effects on the topological surface states.

Ultrafast lattice excitation may drive coherent phonon motion \cite{Zeiger_1992_theory,Merlin_1997_generating}, which by virtue of the electron-phonon interaction, is accompanied by oscillations in the binding energies of electronic states \cite{Khan_1984_deformation}. Coherent phonons in bulk materials are characterized by wavevector $q\approx0$  \cite{Kuznetsov_1994_theory}, and are driven by Raman excitation mechanisms \cite{Merlin_1997_generating,Dekorsy_2000_coherent}.  The femtosecond-scale lattice and electron dynamics can be separately measured by time-resolved x-ray diffraction (trXRD) and time- and angle- resolved photoemission spectroscopy (trARPES), respectively. This combination of techniques provides electronic band, momentum, and phonon mode specificity \cite{Rettig_2015_ultrafast,Gerber_2017_femtosecond}, constituting a powerful toolset for investigating how electronic states respond to coherent lattice motion initiated by ultrafast optical excitation. 

Previous trARPES studies on topological materials (such as \BS{}, \BT{}, and the related materials Bi and Sb) have observed coherent phonons \cite{Papalazarou_2012_coherent,Golias_2016_observation}, with some reporting a contrast between the response of bulk and surface states \cite{Faure_2013_direct,Sobota_2014_distinguishing,Sakamoto_2022_connection}. As this difference could provide a handle for selective control over the surface states, it is important to have a more general understanding of how disparate bulk/surface responses   may be driven by optical excitation.  

Here we report combined trARPES and trXRD measurements of coherent phonons in the prototypical topological insulator \BT{}. We observe  additional   frequencies coupling to the surface states as compared to the bulk states, which are also not observed in bulk-sensitive probes.   This behavior is reproduced in density functional theory calculations, and corroborated by measurements on \BS{}.  We describe how   the photoexcited surface vibrational spectrum can be understood as a continuum of surface-projected bulk modes, including both Raman- and infrared-branches, with possible contributions from surface-localized phonon modes when they exist. This contrasts the expectation from conventional (bulk-sensitive) optical pump-probe experiments, in which only a discrete number of Raman-active modes with $q\approx0$ are observed. Our work offers a comprehensive view of how optical excitations drive coherent motion in the local symmetry-broken environment of the surface, and provides a useful   framework for identifying and manipulating topological states with coherent phonons.

\begin{figure*} 
	\includegraphics[width=\dblcol{}]{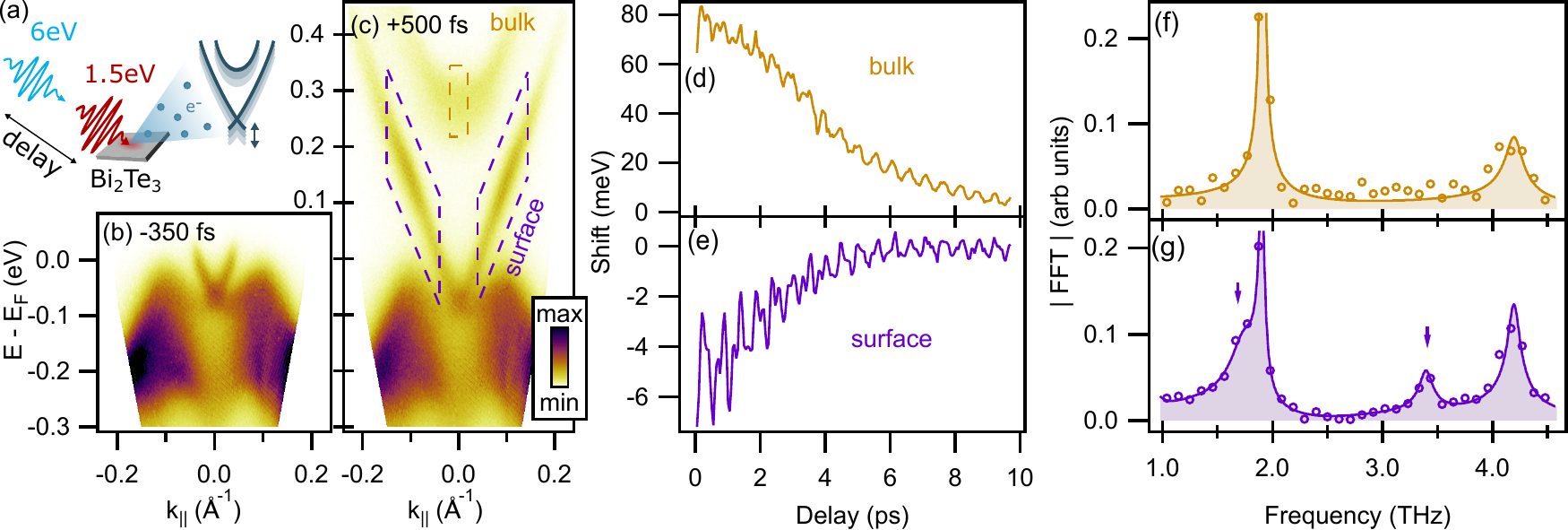}
	\caption{(a) Schematic of time-resolved ARPES measurements on \BT{} to measure electronic binding energy oscillations. (b) ARPES spectrum along the $\overline{\Gamma \textrm{K}}$ direction before and (c) after excitation at the pump-probe delays indicated. The bulk conduction band and most of the surface state are unoccupied in equilibrium, but become partially filled by the excitation. (d) Time-dependent shift in the binding energy of the bulk conduction band and (e) surface state within the  energy-momentum windows indicated by dashed lines in panel (c). (f) Magnitude of the Fourier transform of the bulk conduction band and (g) surface state dynamics after subtracting a slowly varying background ($6^{th}$ order polynomial for the bulk and exponential for the surface). Points are from the data and lines are fits (see Section~\ref{APP_FitMethod} for fitting methodology and Table ~\ref{Table_FitTrARPES} for fit parameters). The arrows highlight two  modes  observed in the surface state but not in the bulk.}
	\label{Fig_ARPES}
\end{figure*}

\section{Methods}

The trARPES system is based on a Ti:sapphire regenerative amplifier outputting 1.5~eV, 35~fs pulses at a repetition rate of 312~kHz. The photon energy of 1.5~eV is used to pump and its fourth harmonic at 6.0~eV is used to probe, as shown in Fig.~\ref{Fig_ARPES}(a). The pump and probe, both $p$-polarized, were focused to spot sizes $80\times 82$ and $47\times 49$~$\mu$m$^2$  full width at half maximum, respectively, at an incident angle 50$^{\circ}$ with respect to normal. The incident pump fluence (not accounting for sample absorption) was 0.36~mJ/cm$^2$, and the time resolution was measured to be 66~fs from a cross-correlation between pump and probe. The energy resolution was $\sim35$~meV. Samples were single-crystals of \BT{} cleaved \emph{in situ}. The sample temperature was set to 27~K, with average heating from the pump leading to an effective measurement temperature of 86~K as estimated from the Fermi-Dirac distribution measured before $t_0$.

\begin{table}
\caption{Fitting parameters for Figs.\ref{Fig_ARPES}(f)-(g)}
\label{Table_FitTrARPES}
\centering
\begin{tabular}{cccc}
\hline\hline
\multicolumn{2}{c}{Bulk} & \multicolumn{2}{c}{Surface}\\
$f$ [THz] & $\tau$ [ps] & $f$ [THz] & $\tau$ [ps] \\
\hline
- & - & $1.73 \pm 0.02$ & $1.3 \pm 0.2$ \\
$1.910\pm0.003$ & $22\pm 9$ & $1.904 \pm 0.004$ & $7 \pm 1$ \\
- & - & $3.40 \pm 0.01$ & $2.7 \pm 0.5$ \\
$4.20\pm0.02$ & $2.1\pm 0.5$ & $4.195 \pm 0.005$ & $2.9 \pm 0.2$ \\
\hline\hline
\end{tabular}
\end{table}

The trXRD measurements were performed at the Linac Coherent Light Source at SLAC National Accelerator Laboratory with 1.5~eV pump ($p$-polarized with incident fluence of 8.2~mJ/cm$^2$) and 9.5~keV probe. The pump and probe were incident at $2^{\circ}$ and $0.5^{\circ}$ with respect to the sample surface, respectively (grazing incidence). The \BT{} sample was a 50~nm film grown by MBE on a BaF$_2$ substrate, with the trigonal axis perpendicular to the surface, and measured at room temperature.

We carried out simulations on a 5 quintuple layer (QL) slab with the ABINIT code \cite{Gonze_2009_ABINIT,Gozne_2016_recent} using the local density approximation \cite{Perdew_1981_self} and the HGH pseudopotentials \cite{Hartwigsen_1998_relativistic}. The interatomic force constants matrix was explicitly obtained on the 5~QL system using DFT ground state calculations, while  atomic forces due to photoexcitation were computed using constrained DFT in the one chemical potential approach \cite{Murray_2005_effect}. The force matrix was then extended to a thicker 500~QL slab to study the behavior of phonons near the surface before calculating the resulting motion in a dynamical matrix formalism (see Section~\ref{APP_CompDetails} for details).   

\section{Results}

The trARPES spectra of \BT{} before and after optical excitation are shown in Figs.~\ref{Fig_ARPES}(b) and (c), respectively. The spectral features can be readily assigned by comparison to conventional ARPES measurements \cite{Chen_2009_experimental}. The spectrum before excitation indicates $p$-type doping due to the Fermi level \EF{} being pinned near the top of the valence band (broad M-shaped band below \EF). The surface state (V-shaped band centered at $\overline{\Gamma}$) is only partially occupied, and the conduction band is completely unoccupied. After excitation, a hot electron distribution extends 100s of meV above \EF{} \cite{Sobota_2012_ultrafast,Wang_2012_measurement}, allowing the conduction band (parabola above $E_{\textrm{F}}+0.2$~eV) to be measured. The time evolution reveals pronounced oscillations in the binding energy of all bands (see movie in \cite{SOM}). We fit the energy distribution curves (EDCs) as a function of delay and parallel momentum $k_{\vert\vert}$. To enhance the signal-to-noise, we average the fit results within the $k_{\vert\vert}$ windows indicated by dashed lines in panel (c). The resulting binding energy shifts are shown in Figs.~\ref{Fig_ARPES}(d) and (e) for the bulk conduction band and surface state, respectively. 

To isolate the oscillatory components, we subtract the slowly varying backgrounds and plot the amplitude of the fast Fourier transforms (FFT) in Figs.~\ref{Fig_ARPES}(f) and (g).  The bulk FFT is well described by a fit consisting of two damped harmonic oscillators, while the surface FFT is fit with two additional peaks (see Section~\ref{APP_FitMethod} for our methodology for fitting the real and imaginary parts of the FFT). The frequencies $f$ and damping time constants $\tau$ from the fits are reported in Table~\ref{Table_FitTrARPES}. Due to the similarity of their frequencies, we identify the two bulk modes near $1.9$~THz and $4.2$~THz with the second and fourth modes observed in the surface state. The same two modes appear in  transient optical reflectivity measurements (Section~\ref{APP_Reflectivity}).

\begin{figure} 
	\includegraphics[width=\sglcol{}]{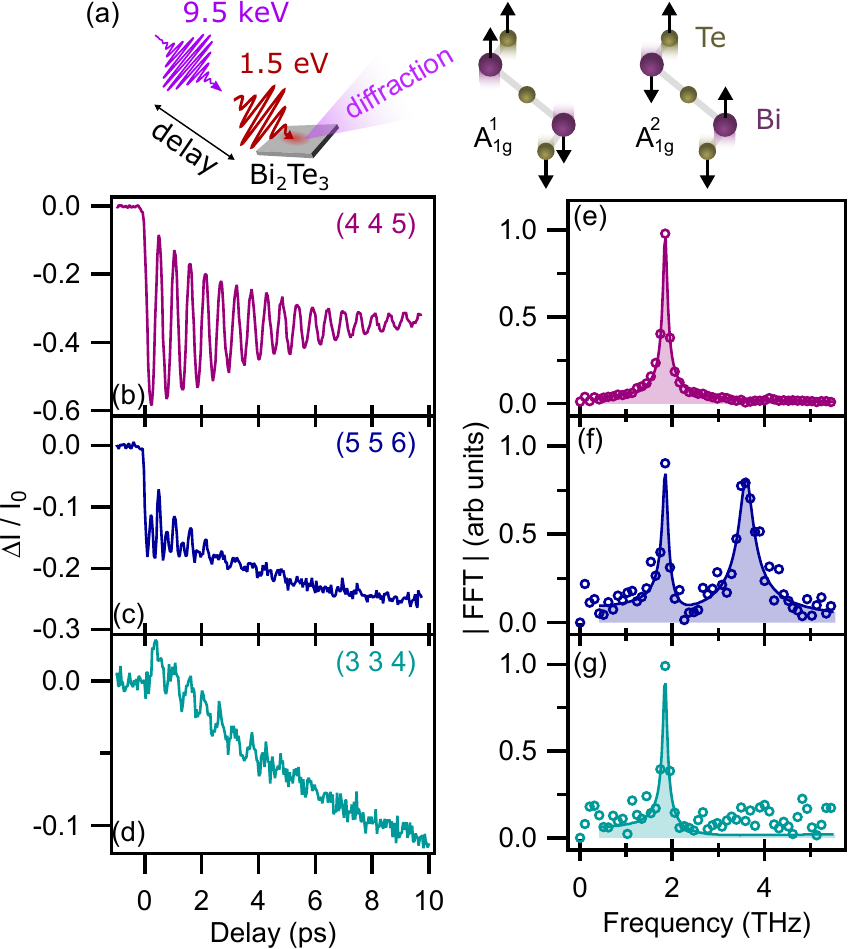}
	\caption{(a) Time-resolved x-ray diffraction measurement to measure coherent lattice dynamics. (b)-(d) Time-dependent relative change in the intensities of the $(h\: k\: l)$ = (4 4 5), (5 5 6), and (3 3 4) Bragg peaks. (e)-(g) The corresponding Fourier transforms after exponential background subtraction. Points are from the data, and solid lines are a fit. Two modes are observed, with $(h+k+l)$-dependent amplitudes consistent with modes of \Ag{1} and \Ag{2} symmetries. 
	}
	\label{Fig_LCLS}
\end{figure}

We used trXRD (Fig.~\ref{Fig_LCLS}(a)) to characterize the lattice distortions associated with these modes. With 5 atoms per rhombohedral unit cell, the optical modes transform as 2\Ag{}+2\Eg{}+2\Au{}+2\Eu{}, with the $\textrm{A}$-modes giving out-of-plane ($c$-axis) displacements and $\textrm{E}$-modes in-plane \cite{Richter_1977_a}.
By measuring 6 distinct Bragg peaks, the measurements fully constrain the eigenvectors describing $c$-axis deformations. Figs.~\ref{Fig_LCLS}(b)-(d) show the transient diffracted intensities of the (4 4 5), (5 5 6), and (3 3 4) Bragg peaks (see Section~\ref{APP_StructFact} for others), with the corresponding FFTs in panels (e)-(g). Two modes are observed at frequencies of 1.85~THz and 3.60~THz, and their contributions are strongly Bragg-peak dependent. The Bragg peak dependence for both modes is well described by distortions of \Ag{1} and \Ag{2} symmetries, as sketched in Fig.~\ref{Fig_LCLS}(a) (see Section~\ref{APP_StructFact} for structure factor analysis).

\begin{figure} 
	\includegraphics[width=\sglcol{}]{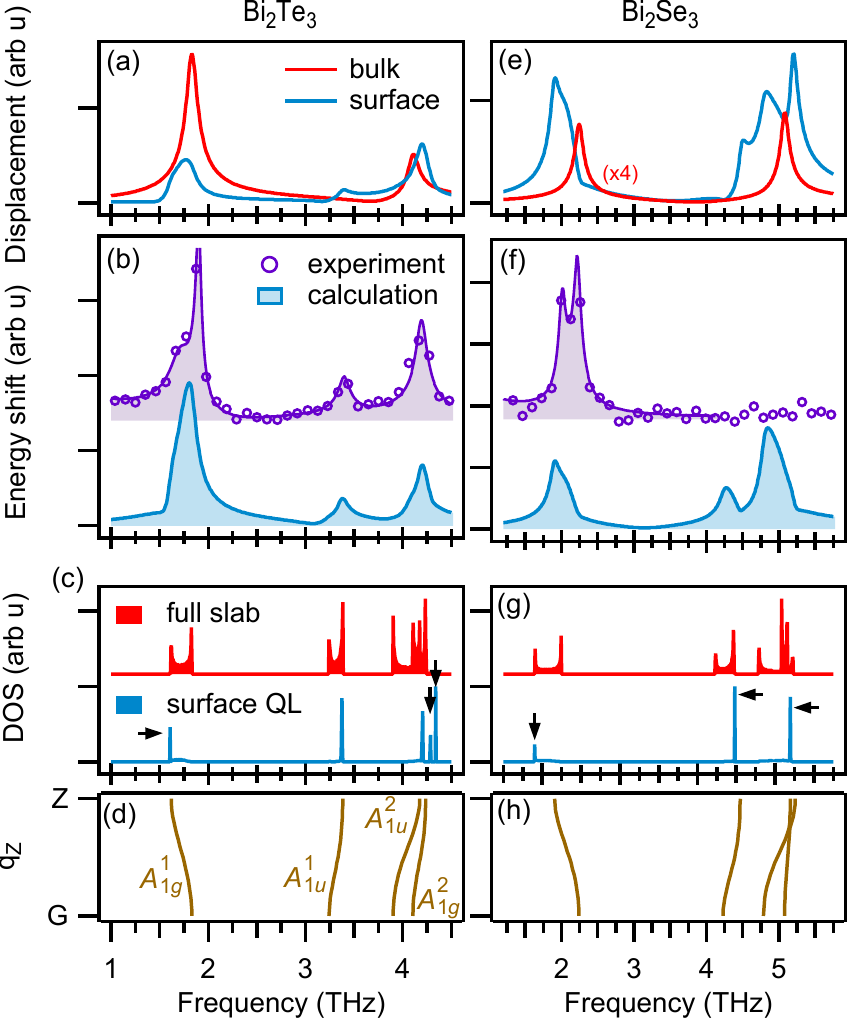}
	\caption{Theoretical calculation of surface and bulk lattice dynamics in \BT{} and \BS{} using DFT and a dynamical matrix formalism. (a) Fourier transforms of the Bi atomic motion at the surface and center (``bulk") of the slab.  (b) The calculated response of the surface state compared to the trARPES data for the surface state (reproduced from Fig.~\ref{Fig_ARPES}(g)), showing overall agreement. (c)   $\overline{\Gamma}$-projected density of states of the full slab and for the top QL only. Arrows denote surface modes outside the bulk continuum. (d) Bulk phonon dispersion curves along $\Gamma$-Z.  
	(e)-(h) Same set of analysis repeated for \BS{}, exhibiting similar phenomenology. Experimental data for \BS{} in (f) reproduced from \cite{Sobota_2014_distinguishing}.}
	\label{Fig_Theory}
\end{figure}

Due to their appearance in bulk-sensitive trXRD measurements, we identify these two modes with those found in the bulk bands measured in trARPES. The fact that the modes measured in trXRD have lower frequencies than those in trARPES is a combined effect of the trXRD measurements being performed at higher temperature and higher fluence, as described in Section~\ref{APP_TempFluenceDep}.   The assignment to \Ag{1} and \Ag{2} modes agrees with the conclusions of previous Raman \cite{Richter_1977_a,Russo_2008_Raman,Kullmann_1984_effect} and transient reflectivity \cite{Misochko_2015_polarization} experiments. The absence of the 1.73 and 3.40~THz modes indicates that those modes are associated with near-surface vibrations.

To gain insight into the disparate surface response, we performed photoexcited slab calculations for \BT{} with out-of-plane lattice degrees-of-freedom. In Fig.~\ref{Fig_Theory}(a) we plot the FFT of the displacement of the Bi atom in the surface quintuple layer (``surface'') and in the center (``bulk'') of the slab. The Bi atoms in the bulk oscillate with frequencies of 1.83~THz and 4.10~THz, with no additional modes observed. In contrast, the Bi atoms at the surface exhibit a broadened response below 1.8~THz, with an additional feature near 3.4~THz. Next we used DFT to compute the time-dependent electronic structure resulting from this lattice motion. We extract the binding energy of the surface state at the $\overline{\Gamma}$-point as a function of time delay, and plot the FFT in Fig.~\ref{Fig_Theory}(b), directly compared to the trARPES measurement. The remarkable agreement demonstrates that our calculations capture the dynamics governing the surface state response. To confirm the reproducibility, we performed the same set of calculations for \BS{} in Figs.~\ref{Fig_Theory}(e)-(f), including a comparison to trARPES results from the literature which observed a splitting of the \Ag{1} peak \cite{Sobota_2014_distinguishing}. Overall agreement is again observed, though higher frequency modes were not observed in Ref.~\cite{Sobota_2014_distinguishing}  due to the poorer time resolution of that experiment combined with a weaker response in \BS{}.

Having shown that DFT calculations capture the lattice dynamics coupled to the surface state, we now seek a deeper explanation. We begin by   comparing to the surface-projected phonon density of states (DOS) at $\overline{\Gamma}$. This is plotted in Fig.~\ref{Fig_Theory}(c) for the full slab and for the surface QL only. For reference, the bulk dispersion and mode assignments are shown in (d). Isolated peaks in the surface DOS signify the existence of surface-localized phonon modes. These modes split off from the bulk branches at the Brillouin zone boundaries and agree with the additional frequencies observed in the surface state; namely, the \Ag{1}-branch at the Z-point and the \Au{1}-branch at the $\Gamma$-point.     The same observations apply for \BS{}, as shown in Fig.~\ref{Fig_Theory}(g)-(h).

\begin{figure*} 
	\includegraphics[width=\dblcol{}]{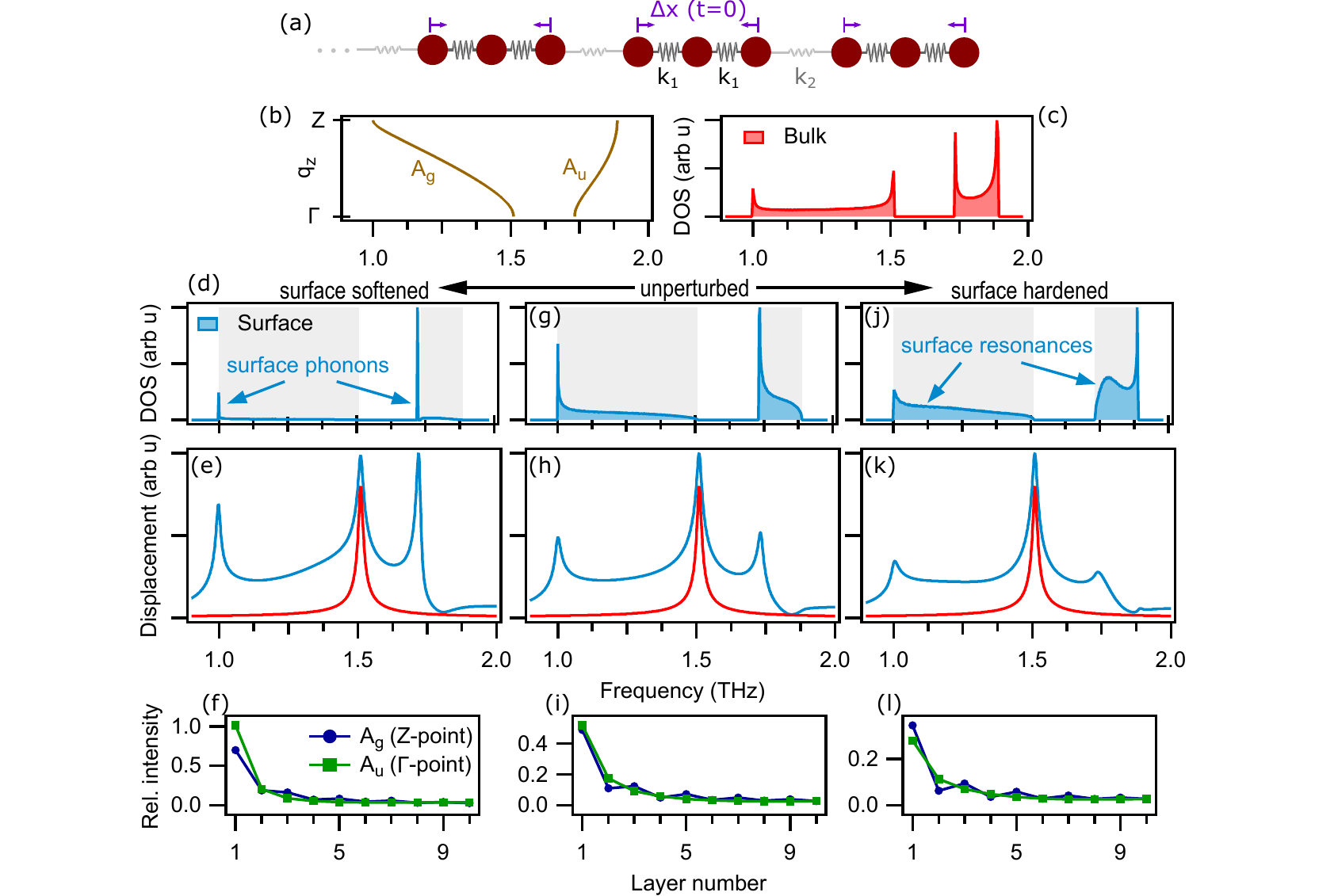}
	\caption{Classical semi-infinite one-dimensional chain of trimers, representing the minimum model containing Raman and infrared phonons. (a) Cartoon of the model. All atoms have identical masses $m$, connected by spring constants $k_1 / m = \left(2\pi\right)^2 \left(1~\textrm{THz} \right)^2$ and $k_2 / m = \left(2\pi\right)^2 \left(0.8~\textrm{THz} \right)^2$. At $t=0$ the trimers are uniformly distorted throughout the chain (see $\Delta x$ arrows) corresponding to excitation of the \Ag{} optical phonon with $q_z=0$.   (b) Phonon dispersion and (c) density of states for the bulk.  (d) Surface density of states and (e) FFT of resulting motion for a bulk (red) and surface (blue) atom for the case in which the surface $k_1$ are softened by 2\%. A damping time of 20~ps is used. (f) The amplitude of the low-frequency and high-frequency peaks relative to the central peak, plotted versus layer number. Corresponding plots for the unperturbed surface are shown in (g)-(i), and for the 2\% hardened surface in (j)-(l).  }
	\label{Fig_Cartoon}
\end{figure*}

  Surface phonons in \BT{} and \BS{} have been identified previously by helium-atom scattering, slab calculations \cite{Zhu_2011_interaction,Ruckhofer_2020_terahertz}, and transient
optical second-harmonic spectroscopy \cite{Bykov_2015_coherent}, so their appearance here is not surprising. However, it is well known that the existence of surface-localized phonons depends sensitively on the details of the surface termination \cite{Kress_1991_surface,Benedek_2020_surface}. Moreover, it is unclear how these modes, particularly those derived from infrared-active branches, are driven by an optical excitation. To examine these subtleties,   we simulate the simplest model system with both Raman and infrared modes: a classical, semi-infinite 1D chain of trimers, as shown in Fig.~\ref{Fig_Cartoon}(a). To mimic photoexcitation,   which must be spatially homogeneous on atomic lengthscales and only couple to Raman-active modes,   the initial conditions are set to symmetrically displace the outer atoms of each trimer with respect to the center one,   corresponding to a $q_z=0$, \Ag{1} mode (see arrows). The bulk dispersion and surface-projected DOS are plotted in   Fig.~\ref{Fig_Cartoon}(b) and (c), respectively.

  We now consider three scenarios: (d) surface softening, in which the spring constants $k_1$ in the top layer are reduced by 2\%, (f) no surface perturbation, and (h) surface hardening, with the surface $k_1$ increased by 2\%. In each case we plot the surface DOS, with the frequency range spanned by the bulk DOS shaded in gray, for reference. This reveals the characteristic sensitivity of surface phonons to surface conditions: in the softened case, localized modes split off from the bulk continuum, whereas in the hardened case, those modes overlap the continuum and form broadened resonances. 

The FFTs of the ensuing atomic motion for an atom at surface (blue) and in the bulk (red) are shown in Figs.~\ref{Fig_Cartoon}(e), (h), and (k) for the three respective scenarios. The bulk spectra are independent of the surface conditions and exhibit a single peak at the frequency of the \Ag{} mode at $q_z=0$, as generally expected for optical excitation. Despite the dramatic difference in surface DOS between these scenarios, the surface atomic motion is qualitative similar, with additional peaks near the Z-point frequency of the \Ag{} branch and $\Gamma$-point frequency of the \Au{} branch. Only the ratio of peak amplitudes, and their precise frequencies, is affected by the degree of surface perturbation. In all cases, these additional peaks drop off rapidly away from the surface (panels (f), (i), and (l)). 
 
\begin{figure} 
	\includegraphics[width=\sglcol{}]{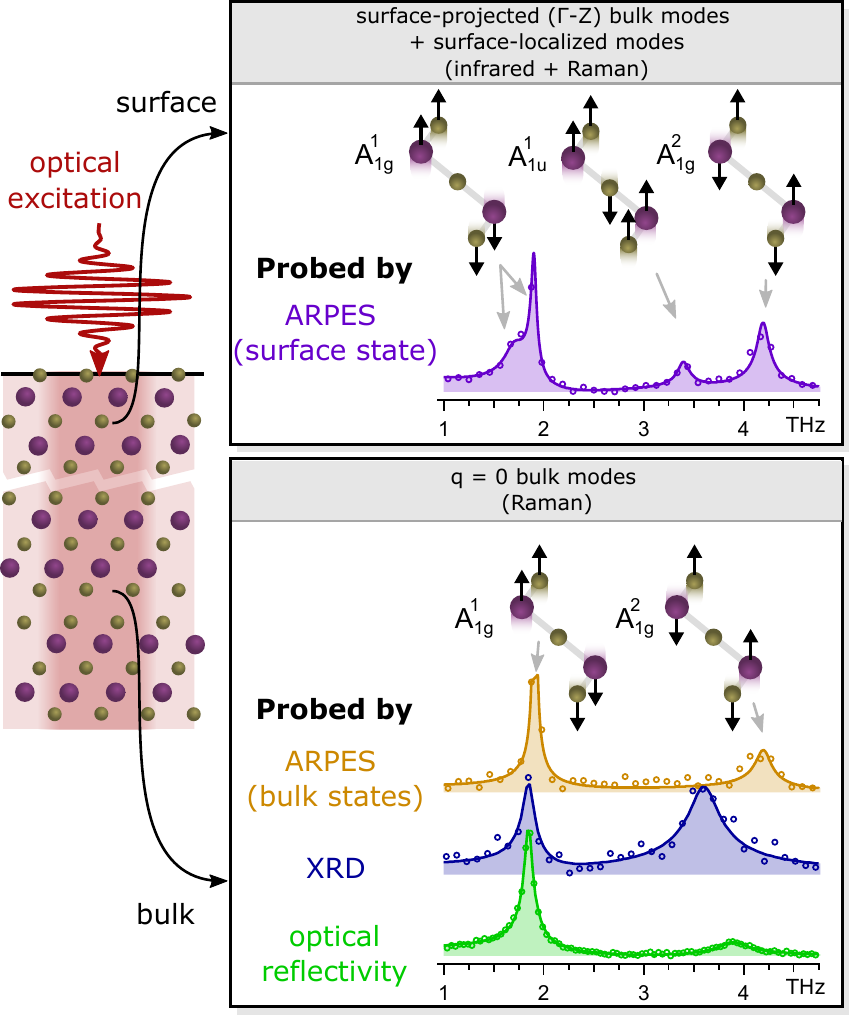}
	\caption{Complementary sensitivities of multi-modal probes. Near the surface (within the top quintuple layer), atomic motion involves  the surface-projected bulk phonon spectrum with infrared- and Raman-active branches in addition to surface-localized modes.   These oscillations are probed by trARPES measurements of the surface state. Deep within the bulk, only Raman-active modes at $q_z=0$ are excited. These are probed by trARPES measurements of the bulk states, as well as trXRD and transient optical reflectivity. The different frequency of the \Ag{2} mode in XRD is due to its different measurement temperature and fluence.  }
	\label{Fig_CartoonTechniques}
\end{figure}

\section{Discussion and Conclusions}

  From our photoexcited slab analysis (Fig.~\ref{Fig_Theory}) and simple model (Fig.~\ref{Fig_Cartoon}), we can draw several conclusions about the behavior of photoexcited surfaces. First, despite the long lengthscale of the excitation, the boundary conditions allow localized vibrations to be driven at the surface. Of course, experimental observation requires a probe sensitive to  near-surface displacements; photoemission from the surface state of a topological insulator is suitable because its wavefunction is largely localized within the top QL \cite{Pertsova_2014_probing}. Second, this motion involves both infrared- and Raman- active branches, despite the Raman-selectivity of the excitation. And finally, our toy model reveals that these observations are independent of the microscopic details of atomic bonding at the surface, which only quantitatively affect the spectrum of the resulting motion. In particular, additional peaks in the surface state spectrum do not imply the existence of split-off surface modes.

It is enlightening to build upon this phenomenological description with a rigorous mathematical foundation.  In Section~\ref{surface_qz_theory} we analytically compute the photoexcited motion of a one-dimensional chain which optionally supports localized surface modes. The main conclusion is that the surface atomic motion is describable as a summation over surface-localized modes (when they exist) plus the surface-projected bulk phonon DOS along $\Gamma$-Z. The latter reflects the fact that each bulk mode is a standing wave which is delocalized across the entire crystal. Therefore, the localized motion of surface atoms must necessarily include a superposition over a continuum of bulk modes.

These principles are well known in helium atom scattering and electron energy loss spectroscopy  \cite{Kress_1991_surface,Benedek_2020_surface}. In these experiments, the colliding particles have appreciable momentum ($\sim1$~$\angstrom^{-1}$) and short range ($<1$~nm) interaction lengths, which allows them to generate localized excitations at all wavevectors. In contrast, the optical excitations employed in an ultrafast experiment have negligible momentum ($\sim10^{-3}$~$\angstrom^{-1}$), are spatially homogeneous over atomic length scales ($>10$~nm penetration depth), and couple to lattice excitations rather indirectly through Raman processes. It is therefore \emph{a priori} not obvious that optical photons should be capable of driving similar lattice excitation spectra, including the full surface-projected bulk DOS and modes of infrared symmetry. These results follow from the generically broken symmetries at a surface: (1) The loss of translational symmetry requires the participation of a broad superposition of bulk modes, and (2) the loss of inversion symmetry locally lifts the distinction between Raman-active and infrared-active modes. 

As mentioned, the existence of localized modes can depend sensitively on microscopic details, and requires careful experimentation and modeling to discern \cite{Zhu_2011_interaction,Ruckhofer_2020_terahertz}. Empirically, it is useful to note that surface-localized phonon frequencies often split only weakly from the bulk continuum, which implies that bulk phonon dispersions can be useful for identification of frequencies in surface coherent phonon data. For   the \Ag{1} branch of \BT{},  neutron scattering   determined bulk frequencies of 1.93~THz and 1.65~THz at $\Gamma$ and Z, respectively \cite{Wagner_1978_lattice}, which compare favorably to the frequencies of 1.90~THz and 1.73~THz observed at the surface here. This picture extends beyond \BS{} and \BT{}: For example, a recent trARPES work detected an additional coherent mode coupling to the surface state of Sb with a frequency 12.7\% higher than its bulk counterpart \cite{Sakamoto_2022_connection}, to be compared with neutron scattering, which measured a $\sim12$\% higher frequency at the zone boundary compared to the $\Gamma$-point.

Returning now to the observation of an infrared-mode, the assignment of the $3.40$~THz peak to an \Au{1} mode has been the subject of some debate in the literature. A number of Raman studies detected modes near this frequency in nanocrystals or thin films, and identified it as the \Au{1} mode becoming Raman-active due to the reduced dimensionality of the samples 
\cite{Teweldebrhan_2010_exfoliation,Souza_2011_structural,Shahil_2010_crystal,Ren_2012_large,Kung_2017_surface,Mal_2019_vibrational,Goyal_2010_mechanically,Wang_2013_in,He_2012_observation}. However, this interpretation has been challenged by Raman studies on bulk \BS{} and \BT{}, where the mode was assigned to a surface phonon \cite{Boulares_2018_surface}. As we discussed above,   the surface coherent phonon spectrum is quite insensitive to microscopic surface details, so our results are agnostic as to whether a localized mode is indeed split off.

As summarized in Fig.~\ref{Fig_CartoonTechniques}, our results highlight the power of multimodal probes, in which the complementary sensitivities of different techniques are exploited to form a comprehensive understanding of an underlying phenomenon.   The concepts invoked here have broadly-applicable predictive power for understanding surface dynamics, and establish a more general framework than previous approaches,   which computed only local energy landscapes rather than consider the bulk/surface phonon structure in entirety   \cite{Sobota_2014_distinguishing}.  For the case of topological insulators specifically, these phenomena offer a practical spectroscopic application: in more controversial materials, where the existence of a surface state is in dispute, we propose that the coherent phonon spectrum may provide a fingerprint by which to unambiguously distinguish surface from bulk states.  From the ``ultrafast control'' point-of-view,  our results imply that surface states can be modulated at frequencies separate from those of the bulk states. This may enable double-pulse coherent control schemes to ``cancel'' the atomic motion in the bulk, while simultaneously allowing the surface atoms to independently oscillate \cite{Hase_1996_optical}. The ability to create surface-localized perturbations should bolster efforts to tailor surface state wavefunctions using ultrafast pulses.

\begin{acknowledgments}
This work was supported by the U.S. Department of Energy, Office of Basic Energy Sciences.
Use of the Linac Coherent Light Source (LCLS), SLAC National Accelerator Laboratory, is supported by the U.S. Department of Energy, Office of Science, Office of Basic Energy Sciences under Contract No. DE-AC02-76SF00515.
The theory work was supported by Science Foundation Ireland and the Department for the Economy Northern Ireland Investigators Programme under Grant Nos. 15/IA/3160 and 12/IA/1601.  We are grateful for the use of computational facilities at the Irish Centre for High-End Computing (ICHEC). We acknowledge illuminating discussions with A.F. Kemper and H. Soifer.
\end{acknowledgments}

\section{Appendix}

\subsection{Computation details}\label{APP_CompDetails}

Density functional theory (DFT)   and constrained DFT (CDFT)   calculations are performed in the local-density approximation (LDA) \cite{Perdew_1981_self}  employing the Hartwigsen-Goedecker-Hutter (HGH)
norm-conserving pseudopotentials \cite{Hartwigsen_1998_relativistic} using the ABINIT code \cite{Gonze_2009_ABINIT,Gozne_2016_recent}.
All calculations are carried out using the experimental lattice parameters given in Table~\ref{table1}. Atomic positions are fully relaxed along the [001] direction while keeping the in-plane lattice parameters fixed to experimental values.    Spin-orbit coupling (SOC) is included in all calculations.   For electronic band structure calculations, Brillouin zone (BZ) integrations are performed on a  12$\times$12$\times$1 Monkhorst-Pack (MP) $\one$-points mesh in the slab calculations and 8$\times$8$\times$8 mesh in the bulk calculations.  An energy cutoff for the plane waves of 15~Ha is used.

\begin{table}[h]
\caption{Bi$_2$Te$_{3}$ and Bi$_2$Se$_{3}$ lattice parameters taken from Ref.~\onlinecite{Nakajima_1963_the}. $a$, $c$ are the hexagonal lattice constants, and $u$,$\nu$ are the internal parameters describing the position of the atoms inside the unit cell.}
\begin{tabular}{|c|c|c|c|c|}
\hline
             & a & c & $u$ & $\nu$ \\ \hline
Bi$_2$Te$_3$ & 4.386 \AA  & 30.497 \AA  &  0.4000    & 0.2095       \\ \hline
Bi$_2$Se$_3$ & 4.143 \AA  & 28.636 \AA  &   0.4008   & 0.2117       \\ \hline
\end{tabular}\label{table1}
\end{table}

In order to simulate the surface state energy in response to longitudinal lattice dynamics in a Bi$_2$X$_3$ (X = Te,Se) slab  after photo-excitation, we build a one-dimensional chain model   in the out-of-plane direction   using the calculated interatomic force constants matrix ($K_{i,j}$) from DFT.    The force constants are   used to construct the dynamical matrix ($D_{i,j}$) given by

\begin{equation}\label{Dij}
    D_{i,j} = \frac{K_{i,j}}{\sqrt{M_{i}M_{j}}}
\end{equation}

\noindent where $i$ and $j$ are the atomic layer index and $M_i$ is the mass of the atom in layer $i$.   DFT calculations are carried out for the 5-QL slab. We construct the dynamical matrix for the 500-QL slab using the dynamical matrix for the 5-QL slab in the following way. First, we neglect the interactions between the QL that are not the nearest neighbors in the 5-QL slab. Then we assume that the interatomic forces of the two top and bottom QL in the 500-QL slab are the same as those in the 5-QL slab, while the force constants for the other QL of the 500-QL slab are the same as those in the middle QL of the 5-QL slab.   By diagonalising  the dynamical matrix    for the 500-QL slab,   we get the normal modes (normalised eigenvectors) and the frequency of the normal modes (square root of the eigenvalues). After the normal modes and frequencies are obtained, we can calculate the longitudinal motion of each atomic layer in the slab following photo-excitation.

To simulate the experimental conditions, we calculate atomic forces induced by a sudden promotion of valence electrons to unoccupied bands. We start with the equilibrium atomic positions for the ground state system, and then we calculate the resulting forces using constrained density functional theory \cite{Murray_2005_effect}, taking that 0.1\% of valence electrons are promoted to conduction band (for bulk calculations) or the surface states (in the case of a slab). First, the atomic forces are computed for the
5-QL slab and then extended to the 500-QL slab. In performing this expansion, the forces for the top two and
bottom two QLs remain unchanged. The forces for
the central QL are taken to be bulk-like and thus applied to the remaining 496 QLs. The resulting motion
of the 500-QL slab is computed   using the   dynamical matrix
formalism , as detailed below . In the CDFT calculations we assume one chemical potential, i.e. electron and hole populations thermalize rapidly according to the Fermi-Dirac distribution. The electronic temperatures corresponding to $n_c = 0.1$\% are summarized in Table \ref{temperatures}.

\begin{table}[]
\caption{Electronic temperatures (in Kelvin) in the one-chemical potential   constrained density functional theory   calculation for the fixed density of photoexcited carriers of $n_c = 0.1$\%.}
\begin{tabular}{|c|c|c|}
\hline
             & Bulk & 5-QL slab \\ \hline
Bi$_2$Te$_3$ &  1610.43    &    1616.46       \\ \hline
Bi$_2$Se$_3$ &   981.48     &    1211.52       \\ \hline
\end{tabular}\label{temperatures}
\end{table}

In the CDFT calculations, the 5-QL system is photo-excited and the change in force on each atomic
layer of the slab is obtained ($F_i$). The equation
of motion for each atomic layer after photo-excitation is given by

\begin{equation}
    M_i{\ddot{x}} = - \sum_{j} {K_{i,j} x_{j} + F_i} \label{eqmotion}
\end{equation}

\noindent where $F_i$ is the force on the atom $i$   in the ground-state position   due to photo-excitation. We transform the equation of motion to scaled coordinates, $u_i = \sqrt{M_i}x_i$, then transform these into the normal mode coordinate representation: $u_i = \sum_{\lambda}{a_{\lambda} u_{i}^{\lambda}}$, where $\lambda$ is the normal mode index and $u_{i}^{\lambda}$
is the (normalised) eigenvector of the dynamical matrix
(in the atomic basis) for mode $\lambda$. The equation of motion in Eq. \ref{eqmotion} becomes

\begin{equation}
    \ddot{u}_{i} = -\sum_{j}{D_{ij}u_{j} + \frac{F_{i}}{\sqrt{M_i}}}
\end{equation}
By taking the inner product of this set of equations with the eigenvector $\textbf{u}^{\lambda} $, we find

\begin{equation}
    \ddot{a}_{\lambda} = -\omega_{\lambda}^{2}a_{\lambda} + f_{\lambda}
\end{equation}

\noindent where $f_{\lambda}$ is the projection of the photoexcited forces on the phonon mode $\lambda$, i.e. $f_{\lambda} = \sum_{i}{u_{i}^{\lambda}\frac{F_{i}}{\sqrt{M_i}}}$.
Using the initial conditions $a_\lambda(t=0) = 0$ and $\dot{a}_{\lambda}(t=0) = 0$, we find that
\begin{equation}
    a_{\lambda}(t) = \frac{f_{\lambda}}{\omega^{2}_{\lambda}} \left( 1-\cos{(\omega_{\lambda}t )} \right)
\end{equation}
  We calculate the motion of each atomic layer as
\begin{equation}
    u_i(t)  = \sum_{\lambda}{a_{\lambda}(t) u_{i}^{\lambda}}
\end{equation}
  Note that to account for phonon lifetime $\tau$, we add a decay of the motion with time: $u_{i}(t) \longrightarrow u_{i}(t) e^{-\frac{t}{\tau}}$.   In our model, the lifetime of all phonon modes is taken to be 3 ps, which is the same order of magnitude as in our experiments (see Table I). Different values of the phonon lifetime do not change our results qualitatively.

 To calculate the motion of the surface state energy as a function of the wavevector $\textbf{k}$, we couple atomic layers to the surface state energy. This is done by moving the atoms in each atomic layer   of the 5-QL slab   and    computing   how
the surface state energy changes   using DFT. In these calculations, we use the vacuum level of the Hartree potential to align the energies of electronic states. To obtain energy changes due to individual atomic motion for the 500-QL slab, we assume that those of the top and bottom three QL are the same as in the 5-QL slab, and that they are zero in the other QL. This is a reasonable assumption since the energy changes in the middle QL of the 5-QL slab are an order of magnitude smaller than those for the outer QL, and since it is expected that atomic motion in the inner layers of thick slabs does not couple with the surface states whose wave functions are confined to a few outer QL only.   The motion of the surface state energy is calculated by

\begin{equation}
    E^{SS}_{\textbf{k}}(t) = \sum_{i} {\frac{dE^{SS}_{\textbf{k}}}{du_{i}}u_{i}(t) }\label{oscillations}
\end{equation}

\noindent where $dE^{SS}_{\textbf{k}}/du_{i}$ is the deformation potential of the surface state calculated using DFT   for the surface state in the vicinity of the $\bar{\Gamma}$ point in the Brillouin zone (1/150 of the distance between the $\bar{\Gamma}$ and M points from the $\bar{\Gamma}$ point). Our results do not change qualitatively if a different point along the $\bar{\Gamma}$-K and $\bar{\Gamma}$-M lines is taken.   

 Finally, the time evolution of the atomic layer displacement $u_i$ and the surface state energy $E^{SS}_{\textbf{k}}$ is Fourier-transformed into the frequency domain over the period of 100 ps with 10$^4$ time steps.    

\subsection{Surface layer excitation in a simplified, nearest-neighbour model}\label{surface_qz_theory}

  In this section, we analytically demonstrate that the motion of an atom at a photoexcited surface is comprised of localized surface modes (when they exist), plus a broad continuum of surface-projected bulk phonons modes.  It is useful to look at a simplified model in which an optical phonon branch is treated in a nearest-neighbour coupling approximation. Let $u_j$ ($j=1,\dots, N$) be the displacement of the local mode in the $j$th layer. The equations of motion for the interior layers are:
$$
\ddot u_j ~=~ -\Omega^2 u_j - \alpha \left(u_{j-1} + u_{j+1} \right)
$$
for $j=2, \dots, N-1$. For the surface layers:  
$$
\ddot u_1 = -[\Omega^2 + \Delta] u_1 - \alpha \; u_2
$$
$$\ddot u_N = -[\Omega^2 + \Delta] u_N - \alpha \; u_{N-1}
$$
where $\Omega$ is the optical mode frequency and $\alpha$ describes the coupling between neighboring layers. The parameter $\Delta$ allows for tuning the near-surface spring constants. This tight-binding formulation should be a good approximation to physical systems whenever an optical phonon branch is decoupled from neighboring branches. It is mathematically equivalent to a  1D chain of dimers with intralayer spring constant $k_1$ and interlayer spring constant $k_2$, which can be seen by equating $\Omega^2 = 2k_1/m+k_2/m$ and $\alpha = k_2/2m$. The change in surface spring constant is $\Delta k_1 = (\alpha+\Delta)/2m$; that is, the surface spring constant is unperturbed for $\Delta = -\alpha$.   

The solutions are standing waves with wave vector $q$, where the displacement in layer $j$ is:
$$
u_{j,q} ~=~  \sqrt{ 2 \over N} \sin(qj + \delta) \cos(\omega t)
$$
Substituting into the equations of motion for interior, we see that the frequency $\omega$ is given by:
\begin{equation}\label{eq::disp_rel}
\omega^2 = \Omega^2 + 2\alpha \cos q 
\end{equation}
The phase shift $\delta$ is determined by matching the equations of motions for the interior and left surface layers, which gives the condition:
 
$$
\tan\delta = \frac{\Delta\sin q}{\alpha -\Delta\cos q}
$$
 
 The boundary condition at the right surface is identical, which gives the quantization condition for allowed wave vectors $q_m$:
  
 $$ q_m = \frac{1}{N}\left(m\pi - 2\delta_m\right), \quad {\rm where~} m~{\rm is~a~positive~integer}. $$

Note that
$$
{d \delta \over dq} ~=~ { ({\alpha/\Delta}) \cos q - 1 \over \left ( {\alpha/\Delta} - \cos q \right)^2 + \sin^2 q }
$$
\noindent is everywhere a well-behaved function, so successive values of $q$ from 0 to $\pi$ can be generated numerically starting from $q_0=0$ by
$$
 q_{m+1} ~=~ q_{m} ~+~ { \pi \over N + 1 + 2 {d\delta \over dq} }
$$
\noindent where ${d\delta \over dq}$ is evaluated at the estimated mid-point, $q = q_{m} + \pi/(2N+2)$, between $q_m$ and $q_{m+1}$.

If $|\Delta| < |\alpha|$ and, therefore, the difference is relatively small between the local mode frequency in the surface layer and that in the bulk layers,  the phase shift $\delta$ can be chosen to be a smooth function of $q$ and lie always, either in the range $(0, \pi/2)$ or in the range $(-\pi/2,0)$, depending on the sign of $\Delta /\alpha$, but never reaching the limit, $\delta = \pm \pi/2$, with $\delta = 0$ at both $q=0$ and $q=\pi$. In this case, there are $N$ distinct allowed values of $q$ lying in the range $(0,\pi)$ and these standing wave solutions include all $N$ normal modes of the system. 
 (For the threshold case, $\Delta = \alpha$, $\delta = (\pi -q)/2$ for $0 < q< \pi$, and for the other threshold, $\Delta = -\alpha$, $\delta = -q/2$ for $0 < q< \pi$.)
 
 On the other hand, if $|\Delta| >|\alpha|$ and the local mode frequency at the surface is significantly different from the bulk, $\tan\delta$ diverges when $\alpha - \Delta \cos q = 0$ and, if $\delta$ is a smooth function of $q$, it must vary over the range $(-\pi,0)$, being equal to $0$ at $q=0$ and equal to $-\pi$ at $q=\pi$. This reduces to $N-2$ the number of allowed values of $q$ that satisfy the boundary conditions, so that, in order to make up the full total of $N$ normal modes, one normal mode must be localised at each surface. The (normalised) mode at the left surface has the form:
 $$
 u_{j,{\rm loc}} ~=~ \left ( { \alpha \over \Delta } \right )^{j-1}  \sqrt{ 1 - \left( \alpha / \Delta \right )^2 } ,
 $$
 where we have assumed that $N \ln\left| \Delta / \alpha \right| >> 1$ and interaction between localised modes on the right and left surfaces can be neglected.
 The frequency $\omega_{\rm loc}$ of the localised surface mode satisfies
 \begin{equation}
 \omega^2_{\rm loc} ~=~ \Omega^2 + \Delta + {\alpha^2 \over \Delta } ~=~ \Omega^2 + \alpha \left( { \Delta \over \alpha} + { \alpha \over \Delta} \right),
 \end{equation}
which, as expected, lies outside the range $\left[ \Omega^2 -2|\alpha|, \Omega^2 + 2 |\alpha| \right]$ of the bulk mode frequencies-squared.

Let us now examine the effect of photoexcitation.   We assume that photoexcitation causes the equilibrium value of the $A_{1g}^1$ coordinate $u$ in all unit cells to shift suddenly by $a$ at time $t=0$. Then we can find the induced motion by projecting the uniform function, $u_j = a$, onto each normal mode $u_{j,m}$:
$$
a_m = a \; \sum_{j=1}^N u_{j,m}
$$
The induced displacement in layer $j$ as a function of time is then:
$$
u_j(t) ~=~ \sum_{m} a_m u_{j,m} \left[1-\cos(\omega_m t) \right]
$$
where $\omega_m$ is the frequency of normal mode $m$.   For the bulk modes, which we label by allowed vectors $q_m$, 
\begin{align*}
a_m ~=~ &a \; \sqrt{ 2 \over N} \sum_{j=1}^N \sin(q_mj+\delta_m ) \\
~=~ &a \; \sqrt{ 2 \over N} { \cos ( \delta_m + q_m/2 ) \over \sin(q_m/2) }, \qquad {\rm for~} m {\rm ~odd},
\end{align*}
and $a_m = 0$ for $m$ even.
For the localised surface mode (when it exists),
$$
a_{\rm loc} ~=~ a \; \sqrt{1 - \left( {\alpha \over \Delta} \right)^2 }\;  \sum_{j=1}^\infty \left ( { \alpha \over \Delta}  \right )^{j-1} ~=~ a \; \sqrt{ 1 + \alpha/\Delta \over 1 - \alpha/\Delta }~.
$$
Thus, the displacement in layer $j$ as a function of time, due to the bulk modes, is:
\begin{align*}
u_j(t) ~=~ &a \; {2 \over N} \sum_{l=1}^{N/2} \sin(q_{2l-1} j + \delta_{2l-1}) \times \\ 
& { \cos ( \delta_{2l-1} + q_{2l-1}/2 ) \over \sin(q_{2l-1}/2) }\left[1-\cos(\omega_{2l-1} t) \right],
\end{align*}
where $\omega_{2l-1}^2 = \Omega^2 + 2 \alpha \cos \left( q_{2l-1} \right)$.
This includes all normal modes when $|\Delta| < |\alpha|$, but when the localised surface mode exists we must add its contribution to the bulk modes:
$$
u_j^{\rm loc}(t) ~=~ a \left ( 1 + { \alpha \over \Delta } \right) \left ( { \alpha \over \Delta } \right ) ^ {j-1} \left[1-\cos(\omega_{\rm loc} t) \right]
$$
Taking the limit $N \rightarrow \infty$, we can replace the summation over $l$ with an integral over frequency:
\begin{align*}
u_j(t) ~=~ {a \over |\alpha| \pi} \; \int_{\sqrt{\Omega^2 - 2|\alpha|}}^{\sqrt{\Omega^2 + 2|\alpha|}} 
    &{ \sin \left ( qj + \delta \right )  \cos \left ( \delta + q/2 \right )  \over \sin q \;  \sin \left ( q/2 \right ) } \times \\
    &\left[1-\cos(\omega t) \right] \omega \; d\omega~,
\end{align*}
where 
$$
q ~=~ \cos^{-1} \left [ { \omega^2 - \Omega^2 \over 2 \alpha } \right ] 
$$
$$
\delta ~=~ \tan^{-1} \left [ { \Delta \sin q \over \alpha - \Delta \cos q } \right ]
$$
In this integral form, it is clear that the power spectrum of oscillations in layer $j$ is the integrand:
\begin{equation}\label{eq:spectrum}
{P_j(\omega) \over a } ~=~  -{1 \over |\alpha| \pi} {\omega  \sin \left ( qj + \delta \right ) \cos \left ( \delta + q/2 \right )  \over \sin q \;  \sin \left ( q/2 \right ) }  
\end{equation}

Eq.~\ref{eq:spectrum} is the main result of this section. In general, it shows that all wavevectors $q$ are required to describe the motion of an atom in a material upon photoexcitation.  It can be cast into a more transparent form by considering the case of an unperturbed surface ($\Delta = -\alpha$), which represents the threshold for splitting off a surface mode. In this case, the mathematics simplify considerably, and Eq.~\ref{eq:spectrum} can be re-written at the surface $(j=1)$ as:  
$$
P_1(\omega) = -{\omega \over |\alpha| \pi} \frac{1}{\sin q}
$$
which is finite for all $q \in (0,\pi)$ and diverges at the zone center $q=0$ and zone edge $q=\pi$.   Moreover, this is  proportional to the phonon DOS ($1/\sin q$) derived from the dispersion relation in Eq.~\ref{eq::disp_rel}. In a three-dimensional crystal, this would correspond to the surface-projected bulk phonon DOS at $\overline{\Gamma}$.   This clearly demonstrates that the motion at the surface comprises frequencies corresponding to bulk phonons throughout the full Brillouin zone. 

In contrast, if we consider the motion deep into the bulk, then the numerator of Eq.~\ref{eq:spectrum} is a rapidly oscillating function with respect to $q$. This may be evaluated by considering the limit:
$$
 \lim_{j\to\infty} \sin \left( q(j-\frac{1}{2} )\right) = \pi q \delta(q)
$$
where $\delta(q)$ is the Dirac delta function centered at $q=0$.   This is a reflection of the principle that only $q=0$ phonons may be photoexcited   in the bulk limit.

\subsection{Fitting methodology}\label{APP_FitMethod}

\begin{figure} 
	\includegraphics[width=\columnwidth{}]{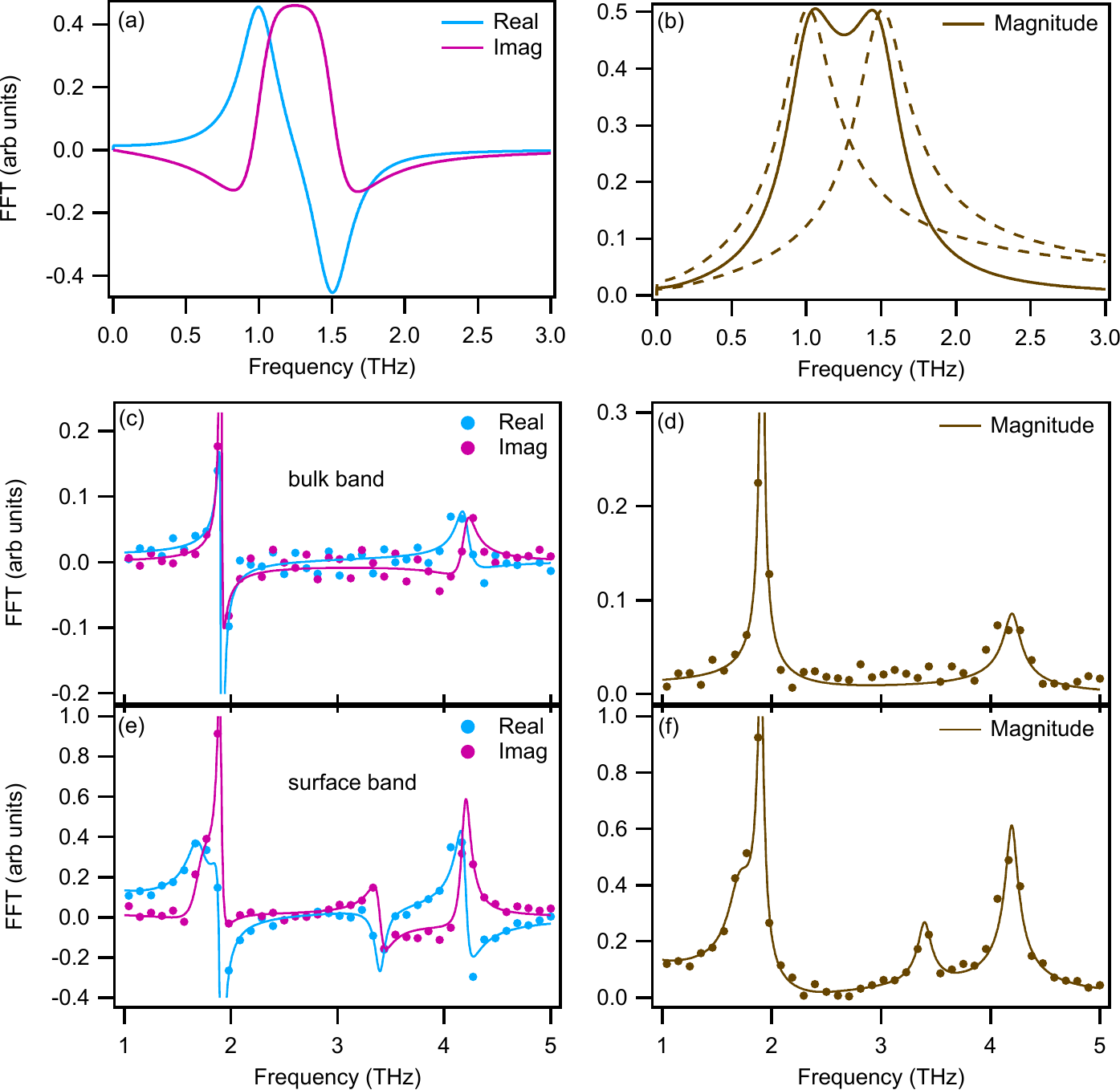}
	\caption{ Overview of the Fourier transform fitting methodology. (a) Real and imaginary parts of the Fourier transform of a simulated data set of two oscillators with equal amplitudes and opposite phases. The frequencies are $f_1=1.0$~THz and $f_2=1.5$~THz with lifetimes $\tau_1 = \tau_2 = 1$~ps. (b) Magnitude of the same Fourier transform. The dashed lines depict the magnitude of each oscillator separately. Note that the total magnitude is not simply the sum of the individual magnitudes due to interference. (c) Simultaneous fit to the real and imaginary parts of the FFT for the bulk band dynamics measured by trARPES. (d) The FFT magnitude resulting from this fit, which is also shown in Fig.~\ref{Fig_ARPES}. (e)-(f) Same analysis but for the surface band.   }
	\label{Fig_FFT_Fit}
\end{figure}

\begin{figure*}
	\includegraphics[width=1.8\columnwidth]{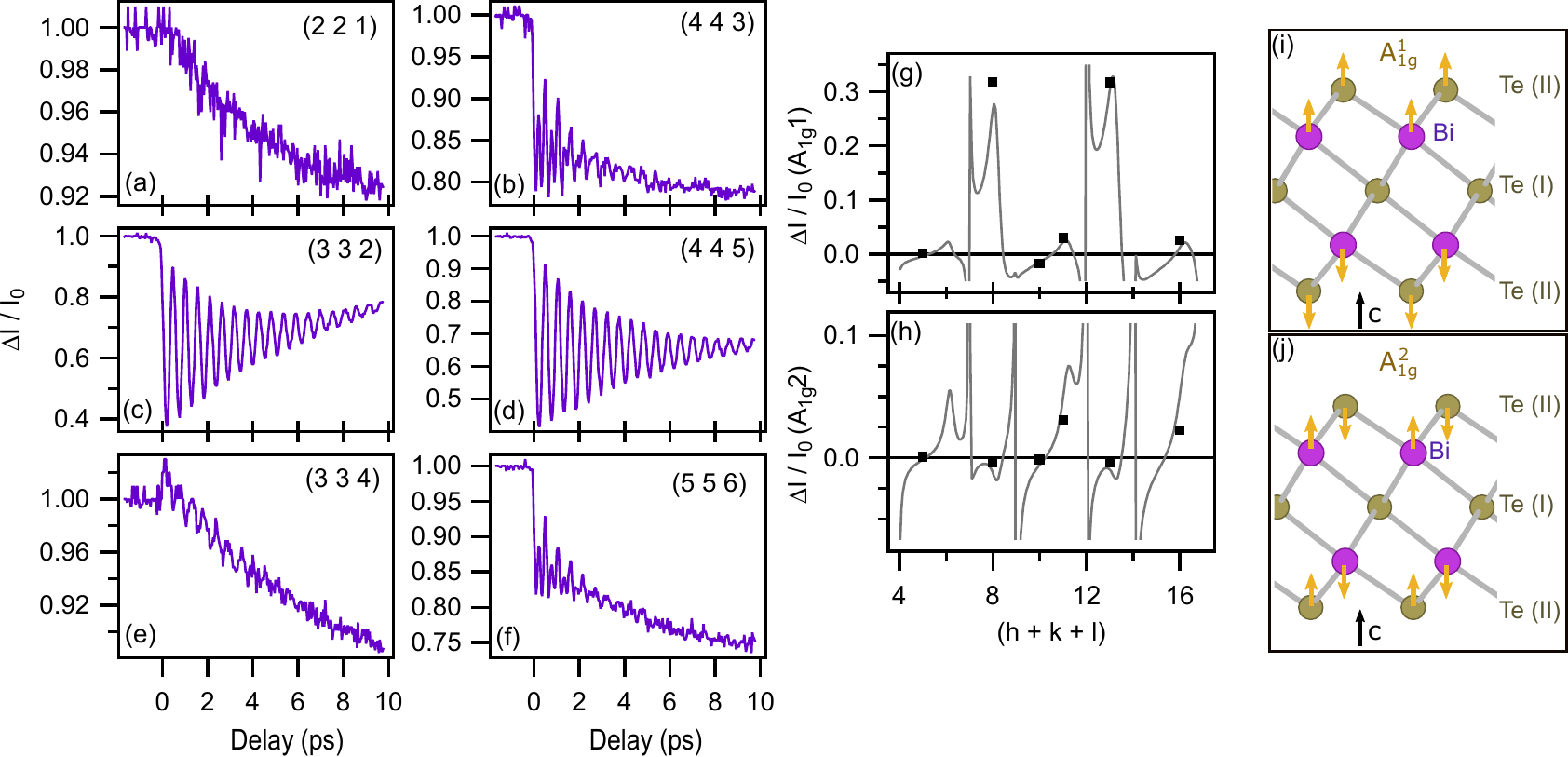}
	\caption{Structure factor analysis of trXRD data. (a)-(f) Time-resolved x-ray diffraction measurement for six Bragg peaks excited by 800~nm with an incident fluence of 8.2~mJ/cm$^2$. 	(g)-(h) Amplitude of both modes extracted by fitting the time-dependent intensities, plotted as a function of $(h+k+l)$ for the six measured peaks. The solid line is a fit from a structure function model assuming modes of \Ag{1} and \Ag{2} symmetries, respectively. This allows for quantitative extraction of the displacements of all five atoms in the unit cell, sketched as arrows in (i) and (j).}
	\label{Fig_LCLS_StrucFac}
\end{figure*}

Oscillatory signals can be analyzed directly in the time or frequency domains. With time domain fitting, it can be difficult to assess the fidelity of the fit, especially when separate modes have overlapping frequencies. On the other hand, frequency domain analysis is typically presented via the \emph{magnitude} of a fast Fourier transform, and thus lacks phase information. Phase information is particularly valuable when separate modes are closely spaced, since interference occurs at overlapping frequencies. To illustrate this point, Figs.~\ref{Fig_FFT_Fit}(a)-(b) show the Fourier transform of a simulated data set consisting of two oscillators. Note that the total magnitude is not simply the sum of the magnitudes for each separate mode.

Motivated by these observations, we perform our analysis in the frequency domain while retaining phase information. The analysis is performed as follows: First, we extract a slowly-varying background from the time-domain data to isolate oscillatory components. Next, we perform a fast Fourier transform while retaining both the real and imaginary parts, thus preserving the phase information. Finally, we perform a \emph{simultaneous} fit to the real and imaginary parts of the Fourier transforms. For the fitting function, we assume the time-domain signal $F(t)$ can be decomposed into a sum of damped oscillators: $F(t) = \sum_j F_j (t)$, where each damped oscillator $F_j(t)$ is given by:

\begin{equation}
F_j(t) = A_j \cos(\omega_j t + \phi_j) e^{-t/\tau_j}
\end{equation}

\noindent where $A_j$, $\omega_j$, $\phi_j$, and $\tau_j$ give the amplitude, frequency, phase, and damping time of the $j$-th oscillator. Then the complex Fourier transform is given by $\hat{F}(\omega) = \sum_j \hat{F}_j (\omega)$ with

\begin{equation}
\hat{F}_j (\omega) =  \frac{\cos(\phi)/\tau_j - \omega_j \sin(\phi) - i \omega \cos(\phi)}{\omega_j^2 - \omega^2 + 1/\tau_j^2 - 2 i \omega/\tau_j}   
\end{equation}

The simultaneous fits using $\operatorname{Re}(\hat{F}_j (\omega) )$ and $\operatorname{Im}(\hat{F}_j (\omega) )$ are shown for the bulk and surface bands in Figs.~\ref{Fig_FFT_Fit}(c) and (e). The corresponding magnitudes, computed \emph{after} the complex fits are performed, are shown in panels (d) and (f). For the bulk band, 2 modes are included in the fit. For the surface band, 5 modes are included. The first mode is held at low frequency ($f \sim 0.1$~THz) and accounts for low-frequency components which remain after background subtraction, and is not regarded as a physical mode.

\subsection{Structure factor analysis}\label{APP_StructFact}

The eigenvectors (atomic displacements) of the two coherent \Ag{} phonons can be determined experimentally by a global fitting of the XRD intensity dynamics.   First, we perform a fit of the time-dependent intensities (Fig.~\ref{Fig_LCLS_StrucFac}(a)-(f)) to a sum of exponentially decaying cosines and a slowly varying background

\begin{equation}
    \frac{\Delta I(\bm{G},t)} {I_0} = \sum_j A_j(\bm{G}) \cos(2\pi f_j t + \phi_j) e^{-\gamma_j t}
    \label{eq:Oscillatory-fit}
\end{equation} 

for each scattering vector $\bm{G}$. Components near zero frequency $f_j$ comprise the slowly varying background, while two frequencies near 2~THz and 4~THz comprise the \Ag{1} and \Ag{2} modes. 

In equilibrium, the intensity of the Bragg peak corresponding to $\bm{G}$ is given by:
\begin{equation}
    I_0(\bm{G}) = \sum_n F_n(|\bm{G}|) \exp\Big({i \big[\bm{G} \cdot \bm{r}_n} \big]\Big)
    \label{static_structure_factor}
\end{equation}  

\noindent where the sum $n$ is taken over the atoms in the unit cell, $F_n(|\bm{G}|)$ is the atomic form factor, and $\bm{r}_n$ is the equilibrium atomic position. 

The time-dependent change due to coherent phonon motion may be written as a sum over normal modes $j$:

\begin{equation}
    I(\bm{G},t) = \sum_{j,n}  F_n(|\bm{G}|) \exp\Big({i \big[\bm{G} \cdot (\bm{r}_n +  u_{j}(t) \bm{\xi}_{j,n})\big]}\Big)
\end{equation} \label{eq:Structure-Factor}

\noindent where $\bm{\xi}_{j,n}$ is the displacement of atom $n$ for normal mode $j$, and $u_{j}(t)$ is the time-dependent motion along normal mode $j$. The fractional intensity change $\Delta I/I_0$ can be computed by dividing by the equilibrium structure factor (Eq.~\ref{static_structure_factor}) and subtracting 1.

However, the symmetry of \BT{} and the fully symmetric \Ag{} modes significantly constrains the symmetry of the problem.  First, all the atoms in the unit cell are stacked along the $z$, or (111) direction, and the \Ag{} modes move the atoms only in the $z$ direction.  Furthermore, the mirror symmetry in this direction and only two atoms reduces this equation to two terms in the sum (Bi and Te atoms).  For each mode $j$ we arrive at a closed-form equation for the time-dependent Bragg peak intensity:
\begin{equation}
\begin{split}
    I_j(\bm{G},t) = 2F_{\textrm{Bi}}(|\bm{G}|) \Big( \cos \big[G_z (z_{\textrm{Bi}} +  u_{j}(t) \xi_{j,\textrm{Bi}})\big] \Big)  +
    \\
    F_{\textrm{Te}}(|\bm{G}|) \Big(1 + 2 \cos \big[G_z (z_{\textrm{Te}} + u_j(t) \xi_{j,\textrm{Te}})\big] \Big)
\end{split}\label{eq:reduced-structure-factor}
\end{equation}

For a given \Ag{} mode, Eq.~\ref{eq:reduced-structure-factor} has only two free parameters: $\xi_{j,\textrm{Bi}}$ and $\xi_{j,\textrm{Te}}$ (all other displacements for Bi and Te atoms are related by symmetry).  Alternatively, one can think of these two free parameters as the Bi/Te atomic displacement ratio (normalized eigenvector) and the total mode amplitude.  Furthermore, because the \Ag{} eigenvectors are orthogonal, the Bi/Te ratio is the same in both eigenvectors, but with opposite sign (Fig.~\ref{Fig_LCLS_StrucFac}(c)).  We measured more than two Bragg peaks for a given fluence, and are thus able to fit the symmetries, eigenvectors, and mode amplitudes simultaneously.

Fig.~\ref{Fig_LCLS_StrucFac}(g) and (h) show the result of our global fit to the eigenvectors and mode intensities for the two \Ag{} modes to the six measured Bragg peaks.  This shows the fractional intensity change $\Delta I(t)/I_0$ as a function of the momentum transfer along the (111) direction $G_z$, represented as h+k+l for the \Ag{1} and \Ag{2} modes in panels (g) and (h), respectively.  The horizontal axis is $G_z$ in reciprocal lattice units (r.l.u).  The solid line in (a) and (b) is a fit of Eq.~\ref{eq:reduced-structure-factor} to our experimental mode amplitudes for each peak, and the points are the mode amplitudes $A_j(\bm{G})$ extracted from a least-squares fit of the the experimental data to Eq.~\ref{eq:Oscillatory-fit}. The periodic structure of the solid lines has a period of five r.l.u because the five atoms in the unit cell are nearly evenly spaced along the (111) direction.  This makes e.g. $(h+k+l) = 10$ not very sensitive to atomic motion, as it is near the maximum of the cosine function in Eq.~\ref{eq:reduced-structure-factor}.

\begin{figure}
\includegraphics[width=0.9\columnwidth{}]{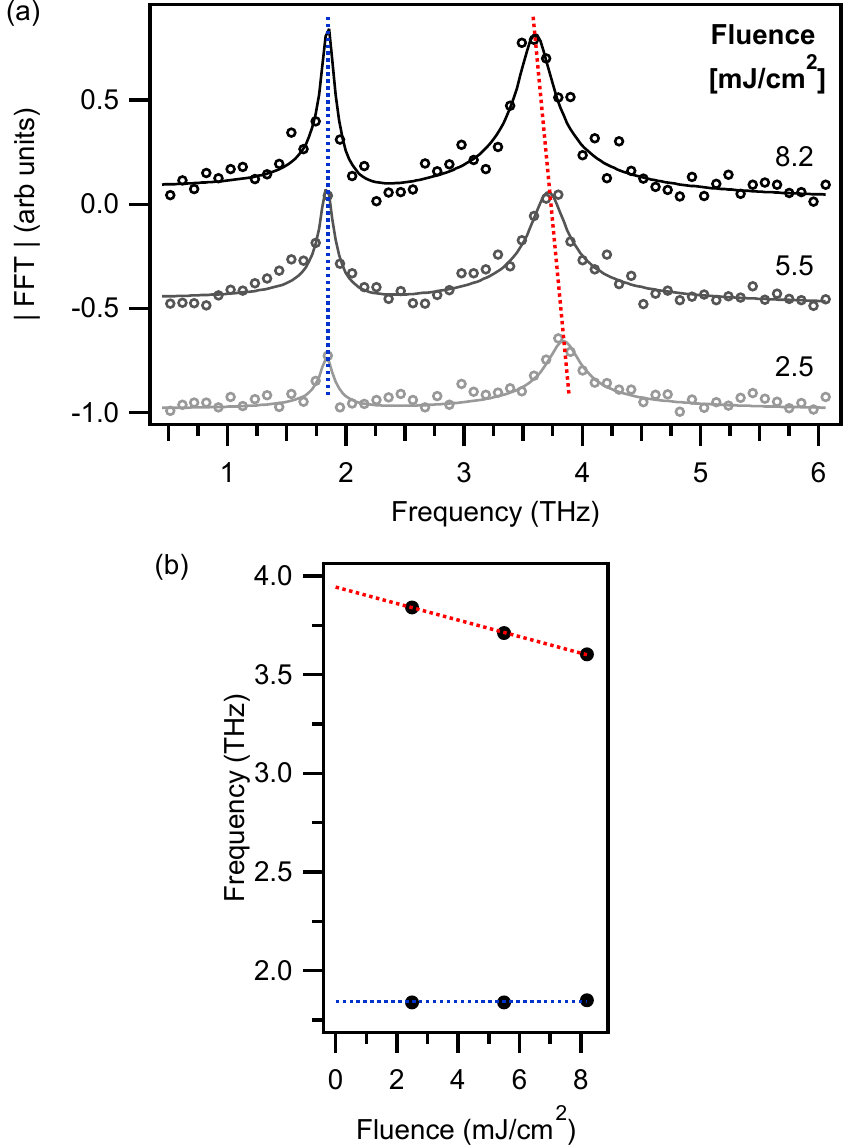}
	\caption{ Fluence-dependence of mode frequencies measured by time-resolved XRD at room temperature. (a) Fourier transform of the (5 5 6) Bragg peak dynamics as a function of incident fluence. Solid curves are fits and dashed lines are guides to the eye. (b) Fluence-dependent frequencies extracted from the fits. The \Ag{1} mode  is fluence-independent at 1.84~THz, while the \Ag{2} mode  extrapolates to $3.94\pm0.05$~THz at zero fluence with a slope of 0.040~THz/(mJ/cm$^2$).   }
	\label{Fig_LCLS_FluenceDep}
\end{figure}

\subsection{Temperature and fluence dependence of mode frequencies}\label{APP_TempFluenceDep}

The reported discrepancy in \Ag{2} frequencies from trXRD and trARPES can be attributed to temperature- and fluence- dependence of its frequency. The temperature-dependence can be estimated from the Raman literature: its frequency was reported to be 4.17~THz at 10~K \cite{Boulares_2018_surface} and 4.02~THz at 300~K \cite{Kullmann_1984_effect}. Interpolating between these values gives 4.16~THz at the trARPES measurement temperature of 27~K, in good agreement with the measured value of 4.20~THz.

The fluence-dependence was measured in the room-temperature trXRD experiment, as shown in Fig.~\ref{Fig_LCLS_FluenceDep}. The frequency of the \Ag{2} mode extrapolated to zero fluence is $3.94\pm0.05$~THz, in reasonable agreement with the value of 4.02~THz reported for room-temperature Raman. This analysis reaffirms the mode assignments made above.

Note that the \Ag{1} frequency exhibits less variation with temperature and fluence.  Raman reported 1.9~THz at 10~K \cite{Boulares_2018_surface} and 1.86~THz at 300~K \cite{Kullmann_1984_effect}. This can be compared with the value of 1.910~THz measured by trARPES at 27~K. Similarly, the room-temperature trXRD measurement shows a fluence-indepndent frequency of 1.84~THz (Fig.~\ref{Fig_LCLS_FluenceDep}).

\subsection{Optical reflectivity measurements}\label{APP_Reflectivity}

\begin{figure}[bh]
	\includegraphics[width=\sglcol{}]{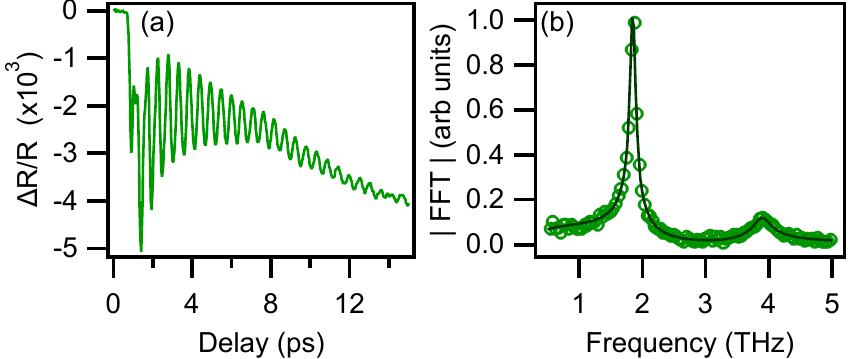}
	\caption{ Time-resolved optical reflectivity measurement on \BT{}. (a) Relative time-dependent change in the reflectivity at 800~nm. (b) Fourier transform after background subtraction. Points are from the data, and solid lines are a fit. Two peaks corresponding to the \Ag{1} and \Ag{2} modes are observed at 1.85~THz and 3.89~THz.  }
	\label{Fig_trR}
\end{figure}

Optical reflectivity measurements were performed at room temperature on Bi$_2$Te$_3$ single crystals (800~nm pump, 800~nm probe), with a 250 kHz amplified Ti:Sapphire laser system at an incident fluence of 1.54~mJ/cm$^2$ (near normal incidence),
shown in Fig.~\ref{Fig_trR}. These measurements were found to be sensitive to the \Ag{1} and \Ag{2} modes only.

%\bibliography{BT_Bib}

\end{document}